# Many-body and correlation effects on parametric polariton amplification in semiconductor microcavities

S. Savasta, O. Di Stefano & R. Girlanda

*INFM and Dipartimento di Fisica della Materia e Tecnologie Fisiche Avanzate, Università di Messina, Salita Sperone 31, I-98166 Messina, Italy*

**Very efficient amplification of light-matter waves (polaritons), that are a superposition of cavity photons and excitons[1] has recently been reported[2-11]. The optical gain curve versus the pump power shows a threshold and then saturates to a maximum value[7,11]. Very recently it has been shown that this limit-value of gain can be greatly enhanced by increasing the exciton-photon coupling rate, allowing to approach room temperature operation[11]. This anomalous enhancement is in contrast with results from present theories[12,13] describing the process. Here we clarify the mechanisms determining gain saturation and explain the observed giant amplification. We show that this enhancement origins from the non-instantaneous nature of exciton-exciton collisions in semiconductors[14] due to many-body correlations. We find that the exciton-photon coupling is able to alter the exciton dynamics during collisions and hence to modify the coupling mechanism at the basis of amplification. These results give precise indications to favour room temperature operation for the realization of all-optical microscopic switches and amplifiers and demonstrate that exciton-exciton collisions in semiconductors can be controlled and engineered.**

Cavity-polaritons are two-dimensional eigenstates of semiconductor microcavities which result from the strong resonant coupling between cavity-photon modes and excitons in embedded quantum wells. The dynamics and hence the resulting energy bands of these mixed quasiparticles are highly distorted respect to those of bare excitons and cavity photons (Fig. 1). The energy dispersion-relations of polaritons anticross when the energy difference between exciton and photon modes is varied. The exciton-photon coupling rate $V$ determines the splitting $2V$ between the two polariton energy bands. Parametric amplification of these mixed light-matter waves able to produce very efficient light amplification has been demonstrated. Coherent amplification of polaritons requires a coupling mechanism able to transfer polaritons from a reservoir (in this case provided by polaritons resonantly excited by a pump laser pulse on the lower polariton dispersion) to the signal mode, while conserving energy and momentum. Polaritons of different modes are coupled via their excitonic content, the coupling being provided by the Coulomb interaction between excitons. The microscopic theory of parametric polariton amplification[12] shows that the resulting coupling strength is given by the exciton-exciton scattering rate $V_{xx} \simeq 6.08 E_b a_0^2$ ($E_b$ is the exciton binding energy and $a_0$ is the two-dimensional exciton Bohr radius) times the exciton fraction of the interacting polariton modes. This effective polariton-polariton interaction scatters a pair of pump polaritons into the lowest-energy state and into a higher-energy state (usually known as the idler mode) (energy and momentum conservation requires $2\omega_k = \omega_0 + \omega_{2k}$, where $\omega_k$ is the energy of a pump polariton injected with an in-plane wavevector **k**) as shown in Fig. 1. The scattering process is stimulated by a weak signal beam injected perpendicular to the cavity (k=0) that is thus greatly amplified. Recently it has been shown that, increasing the exciton-photon coupling $V$ (inserting a large number of quantum wells into the cavity), greatly increases amplification[11] (gains up to 5,000 have been observed), allowing the design of several possible microscopic devices approaching room temperature operation. This spectacular enhancement of gain contrasts with the results of present theories[12,13] describing the process. According to these descriptions, the effective polariton-polariton interaction does not depend on the polariton splitting (2*V*), being the exciton content of the 3 interacting modes independent on *V*. From this observation and from a direct inspection of the pertinent equations, it descends that, increasing the exciton-photon coupling, should not increase the amplification of a weak signal beam in contrast with the spectacular experimental observations [11].

Here we show that the observed great enhancement of gain when increasing the polariton splitting origins from the unique interplay of the non-instantaneous nature of exciton-exciton interactions[14] with the strong coupling regime giving rise to polaritons. The mean-field theory of the optical response of electron-hole pairs assumes that collisions are instantaneous and only the mean free time between collisions matters. Although the mean-field theory has enjoyed considerable success[15], recent studies have revealed that collisions between excitons are more complex and involve multiparticle (at least 4-particle) entangled states, that is a coherent superposition of eigenstates unknown in classical physics[14]. This complexity induced by the Coulomb interaction between electrons, determines the finite duration of collisions[16-20]. We find that the strong coupling of excitons with cavity-photons, giving rise to polaritons, alters the excitonic dynamics during exciton-exciton collisions, producing a modification of the effective scattering rates. Since the dynamics of the coupled excitons and photons is determined by the exciton-photon coupling rate *V*, its variation is able to affect significantly exciton-exciton collisions and hence the amplification process. Thus we show that the possibility of observing very large gains is a consequence of the coherent control that the exciton-photon interaction exerts on the non-instantaneous exciton-exciton collisions. This control able to produce almost decoherence-free exciton-exciton collisions and hence giant parametric amplification attributes a new technological role to many-body and correlation effects in semiconductor.

The time evolution of the coupled exciton ($P_\mathbf{k}$) and photon waves ($E_\mathbf{k}$) including finite duration of exciton-exciton collisions[19,20] can be described by the following set of coupled equations[15],

$$\frac{\partial}{\partial t} E_\mathbf{k} = -(\gamma_c + i\omega_\mathbf{k}^c)E_\mathbf{k} + iVP_\mathbf{k} + t_c E_\mathbf{k}^{in}$$

$$\frac{\partial}{\partial t} P_\mathbf{k} = -(\gamma_x + i\omega_x)P_\mathbf{k} + iVE_\mathbf{k} - i\Omega_\mathbf{k}^{NL}$$

(1)

where $\omega_k$, $\omega_x$, and $\gamma_k$, $\gamma_x$ are the energies and dephasing rates of cavity photons and quantum well excitons respectively. $E_\mathbf{k}^{in}$ describes input light pulses, $t_c$ determines the beam fraction passing the cavity mirror. The intracavity photon-field and the exciton field of a given mode $\mathbf{k}$ are coupled by $V$. The polariton splitting in microcavities can be increased by inserting a large number of quantum wells into the cavity. ($V = V_1\sqrt{N_{eff}}$, where $V_1$ is the exciton-photon coupling for 1 quantum well and the effective number of quantum wells $N_{eff}$ depends on the number of wells inside the cavity and their spatial overlap with the cavity-mode). The relevant nonlinear source term coupling waves of different $\mathbf{k}$ modes is given by $\Omega_\mathbf{k}^{NL} = (\Omega_\mathbf{k}^{sat} + \Omega_\mathbf{k}^{xx})/N_{eff}$, where the first term originates from the phase-space filling of the exciton transition

$$\Omega_\mathbf{k}^{sat} = \frac{V}{n_{sat}} \sum_{\mathbf{k}',\mathbf{k}''} P_\mathbf{q}^* P_{\mathbf{k}''} E_{\mathbf{k}'}$$

(2)

being $n_{sat} = 7/(16\pi a_0^2)$ the exciton saturation density ($\mathbf{k}'$, $\mathbf{k}''$, and $\mathbf{q}$ are tied by the momentum conservation relation $\mathbf{k}+\mathbf{q} = \mathbf{k}'+\mathbf{k}''$). $\Omega_\mathbf{k}^{xx}$ is the Coulomb interaction term. It dominates the coherent polariton-polariton coupling and for co-circularly polarized waves can be written as

$$\Omega_\mathbf{k}^{xx} = \sum_{\mathbf{k}',\mathbf{k}''} V_{xx} P_\mathbf{q}^*(t) P_{\mathbf{k}''}(t) P_{\mathbf{k}'}(t) - iP_\mathbf{q}^*(t) \int_{-\infty}^{t} F(t-t') P_{\mathbf{k}'}(t') P_{\mathbf{k}''}(t') .$$

(3)

$\Omega_{\mathbf{k}}^{xx}$ includes the instantaneous mean-field exciton-exciton interaction term ($V_{xx}$,) plus a non istantaneous interaction term originating from 4-particle correlations. This coherent memory can be interpreted as a non-markovian process involving the 2-particle (excitons) polarization waves interacting with a bath of 4-particle correlations[14,16]. The strong exciton-photon coupling does not modify the memory kernel $F(\tau)$ as a consequence of the fact that 4-particle correlations do not couple to cavity photons, but it is able to alter the phase dynamics of the 2-particle polarization waves $P_{\mathbf{k}}$, during collisions, i.e. on a timescale shorter than the decay time of the memory kernel $F(\tau)$. In this way the exciton-photon coupling $V$ affects the exciton-exciton collisions that govern the polariton amplification process. To better understand this mechanism, let us consider a situation where the pump, the signal and idler energies are all close to the corresponding polariton resonance values and the broadening are small compared to the polariton splitting $2V$. Then it is a good approximation to replace the integral in Eq.(3) with a simpler expression, adopting the Weisskopf-Wigner approximation used to analyse the spontaneous emission between two atomic levels. Within this approximation, the dominant exciton-exciton interaction term for the signal (0) and idler (2**k**) modes can be written as

$$\Omega_{0(2\mathbf{k})}^{b} = \mathcal{F}(\omega_{0(2\mathbf{k})} + \omega_{\mathbf{k}}) | P_{\mathbf{k}} |^2 P_{0(2\mathbf{k})} + \mathcal{F}(2\omega_{\mathbf{k}}) P_{2\mathbf{k}(0)}^{*} P_{\mathbf{k}}^{2}, \qquad (4)$$

where the complex quantity $\mathcal{F}(\omega)$ is the Fourier transform of the memory kernel $F(\tau)$ plus the frequency independent contribution $V_{xx}$ (from the instantaneous interaction term). The first term produces a blue-shift of the polariton resonance and introduces an intensity-dependent dephasing mechanism. They are both proportional to the coherent exciton density $| P_{\mathbf{k}} |^2$ generated by the pump beam. The second term provides the coupling mechanism able to transfer polaritons from the pump to the signal and idler modes. An analogous expression can be derived for the exciton-exciton interaction of the pump mode. This Weisskopf-Wigner description of 4-particle correlation effects in semiconductor microcavities sets out the relevance of polariton pairs in the scattering process. It is the energy of these pairs that determines the effective scattering rates as shown by Eq. (4) and by the corresponding Feynman diagrams (panel b and c of Fig.2). The spectral function $\mathcal{F}(\omega)$ can be calculated by numerical diagonalization of the semiconductor Hamiltonian in the 4-particle subspace. We have calulated it for quantum well excitons following a recent microscopic approach[17,18] based on the T-matrix (Fig. 2a). The obtained 4-particle spectral density displays strong variations within the spectral region of interest around $2\omega_0$. In particular, moving towards the low energy region, the dispersive part Re($\mathcal{F}$) increases while the absorptive part Im($\mathcal{F}$) that contrasts gain goes to zero. We observe that the energies of the pump, signal, and idler polaritons are lowered with the increase of polariton splitting (see Fig.1a). The increase of the exciton-photon coupling $V$, thus is expected to favour the amplification process. This analysis explains at least qualitatively the large increase of gain observed when maximizing the exciton-photon coupling[11]. Moreover it shows that in semiconductor microcavities it is possible to produce almost decoherence-free polariton-polariton collisions on the lower polariton curve. On the contrary collision-induced decoherence increases for light-matter waves on the upper polariton curve.

The gain curves, obtained numerically solving the system of integro-differential equations (1) for GaAlAs-based samples with splitting $2V_1$= 10.6 meV and $2V_2$=15 meV (Fig.3a) fully confirm this analysis. We observe that, an increase of less then a factor 1/3 in the polariton splitting produces an increase of the maximum achievable gain at T = 10K (Fig3**a**) of more than one order of magnitude (Fig.3a). The power dependence of gain shows an almost exponential growth and then saturates at high powers. The saturation of gain is mainly determined by the nonlinear absorption that is most relevant for the idler beam. We observe that the increase of the nonlinear absorbance $\text{Im}[\mathcal{F}(\omega_{2\mathbf{k}} + \omega_{\mathbf{k}})] | P_{\mathbf{k}} |^2$ is highly superlinear because the increase of the pump power produces both a direct increase of the exciton density $| P_{\mathbf{k}} |^2$ and an increase of $\text{Im}\,\mathcal{F}(\omega_{2k}+\omega_k)$ as a consequence of the blue shift the of polariton-pair resonance $\omega_{2k}+\omega_k$ produced by $\text{Re}[\mathcal{F}]\,|P_k|^2$. We observe that mean-field calculations (also reported in Fig. 3a) overestimate the experimentally observed gain of more than 4 order of magnitude, neither are able to reproduce the strong dependence of gain on the exciton-photon coupling *V*, that is direct consequence of the inner dynamics of exciton-exciton collisions. A so dramatic failure of the mean field theory is surprising and clearly shows the technological relevance that many-body correlations and their optical control may have in the future development of semiconductor all-optical devices. Fig.3b displaying results for T=77K, shows how the increase of the exciton-photon coupling favours high-temperature operation. At T=77K the sample $V_1$ is still able to sustain a significant gain about six times larger than that of sample $V_2$.

The low-temperature time dynamics of the parametric amplification is displayed in Fig.4. The appearance of an additional light beam in a direction allowing momentum conservation, also observed experimentally[7,9], confirms the coherent nature of this scattering process involving pairs of polariton waves. Analogous results (not shown) but with a faster decay of the signals, due to the increased exciton decay rate, are found at higher temperature.

The wave-like nature of excitations in solids tells us that collisions between excitons are nothing but interference phenomena. The strong-coupling regime of semiconductor microcavities, altering the phase of the interacting waves provides a means to control the interference process. This coherent control of exciton-exciton interactions is able to produce highly desirable almost decoherence-free collisions that make possible to reach very-high degree of amplification and thus to approach room temperature operation. Almost decoherence-free exciton-exciton interactions and hence structures with large polariton splitting are expected to be essential to reach room temperature operation. Moreover the availability of decoherence-free collisions appears crucial for the quantum control and manipulation of the polariton wavefunction inside the cavity. Understanding that coherent exciton-exciton collisions in semiconductor quantum structures can be controlled opens new perspectives for the realization of entangled collective polariton states for quantum information and computation.

<ack> Style tag for the Acknowledgements.

**Correspondence and requests for materials should be addressed to S.S. (e-mail: savasta@ortica.unime.it).**

**Figure 1** Sketch of the polariton energy versus the in-plane wavevector (**a**), for two different exciton-photon coupling ($V_1$ = 6.4 and $V_2$ = 15 meV). A polariton mode of given wavevector $k$ can be excited by resonant light with incidence angle $\theta$ given by the relation $k = (\omega/c)\sin\theta$. UP and LP denote the splitted upper polariton and lower polariton curves. The parametric polariton-polariton scattering process is also depicted. **b**, Schematic of the experiment.

**Figure 2** Energy dependence of the effective exciton-exciton scattering potential. It is the sum of an energy independent term (the mean field contribution) plus a term originating from 4-particle (2 electrons and 2 holes) states of the quantum well. It has been calculated (following the microscopic approach described in Ref. 15) for a GaAs quantum well 7 nm wide with exciton

binding energy 13.5 meV and using as dephasing rate of the 4-particle states $\Gamma = 2\gamma_x$ = 0.58 meV (corresponding to a temperature T = 10K). The tail of Im[$F(\omega)$] at negative detuning ($\omega < 2\omega_0$) is produced by $\Gamma$ and vanishes for $\Gamma \to 0$. The inset shows the dynamics of the memory kernel $F(\tau)$. **b, c**, Feynman diagrams describing the effective exciton-exciton interaction as modified by polariton effects. **b**, diagram of the interaction determining the parametric scattering of two pump polaritons into a probe and idler polariton. Note that the energy of the final polariton pair is equal to that of the starting pair ($2\omega_k = \omega_0 + \omega_{2k}$) and the strength of the interaction depends on this energy value. **c**, diagram describing self-interaction of polariton pairs producing blue-shift ($\propto \text{Re}\mathcal{F}$) and nonlinear absorption ($\propto \text{Im}\mathcal{F}$). This last diagram is relative to the probe mode, other two analogous diagrams can be drawn for the pump and idler modes.

**Figure 3** The power dependence of the integrated gain (the total light intensity transmitted in the signal direction divided by the intensity transmitted in the absence of the pump beam) calculated for two GaAlAs based samples with splitting $2V_1$ = 10.6 meV (corresponding to $N_{eff}$ = 4) and $2V_2$ = 15 meV ($N_{eff}$ = 8). $I_0 = 10^{13}$ photons per cm$^2$ per pulse. The exciting light pulses have 250 fs duration. The material parameters of the two samples are coincident except for $N_{eff}$. The decay rate of cavity photons through the mirrors is $\gamma_c$ = 0.25 meV. The 7 nm wide quantum wells have a binding energy $E_b$ = 13.5 nm. The homogeneous exciton broadening mainly due to exciton-phonon scattering depends on temperature. Calculations have been performed by using broadenings extracted from linewidth measurements, $2\gamma_x$ = 0.6 meV at T=10K (**a**), and $2\gamma_x$ = 1.4 meV at T=77K (**b**). The gain curves have been obtained fully optimizing incidence angles, central frequencies of the input light beams and cavity mode energy. Maximum gain occurs when the cavity mode energy is lower then the exciton energy by slightly more than V.

**Figure 4** Time-resolved transmission of the probe (with and without the presence of the pump beam) and additional emission (appearing in the presence of the pump beam) along the direction of the idler mode ($\theta \approx 47°$), showing the ultrafast dynamics of the parametric amplification. The sample is $V_1$, the temperature is 10K, and the pump intensity is $I=I_0$. The exciting light pulses have 250 fs duration.

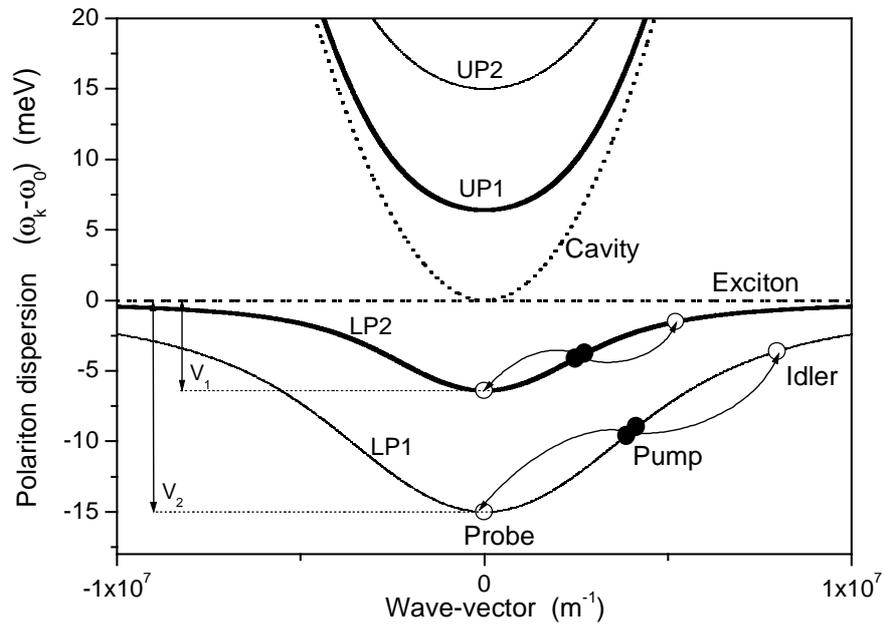

Figure 1a

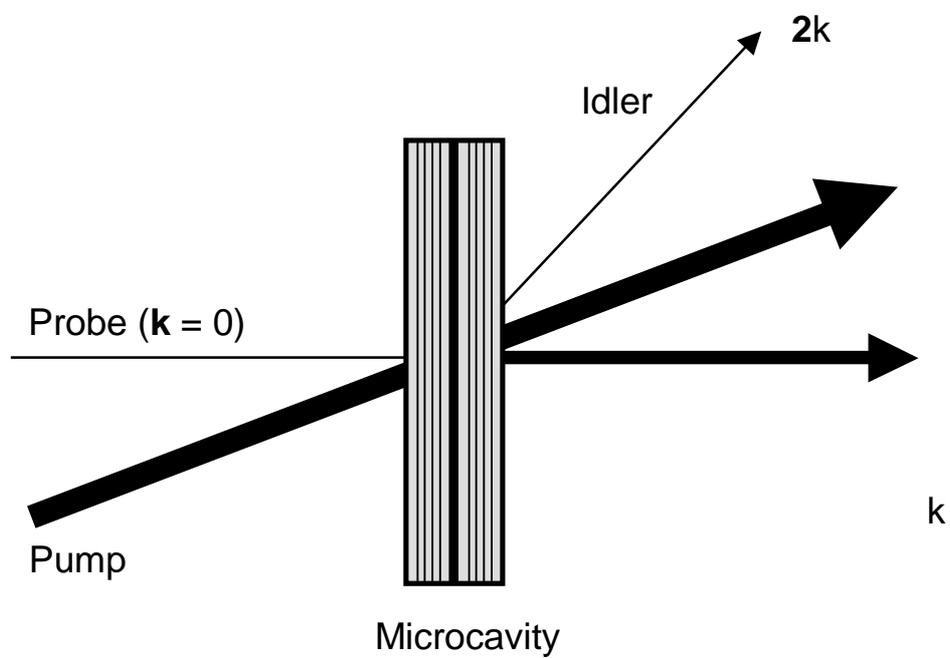

Figure 1b

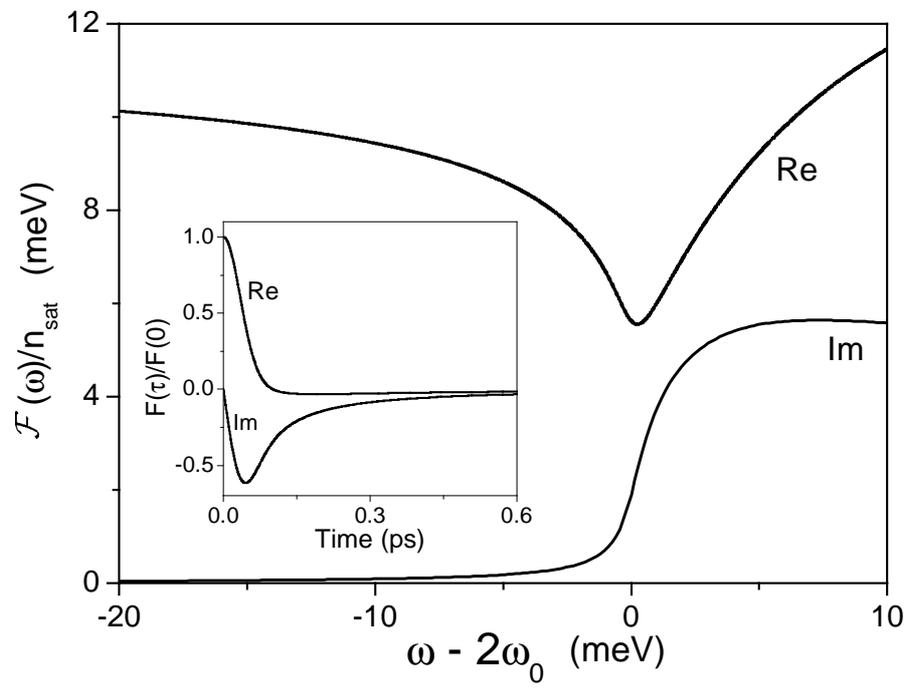

Figure 2a

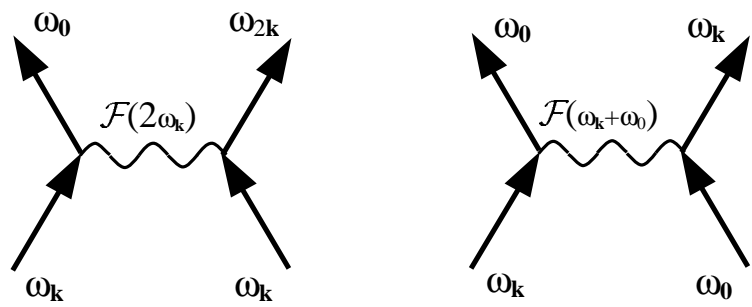

Figure 2b

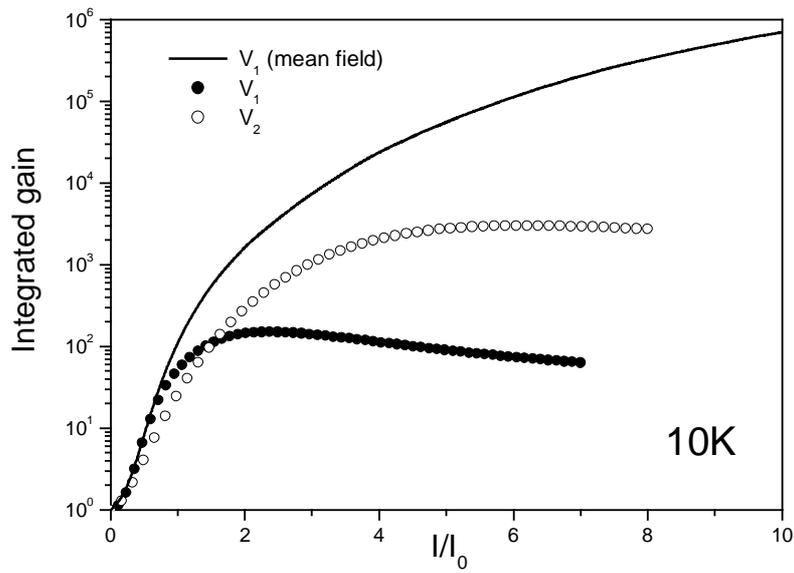

Figure 3a

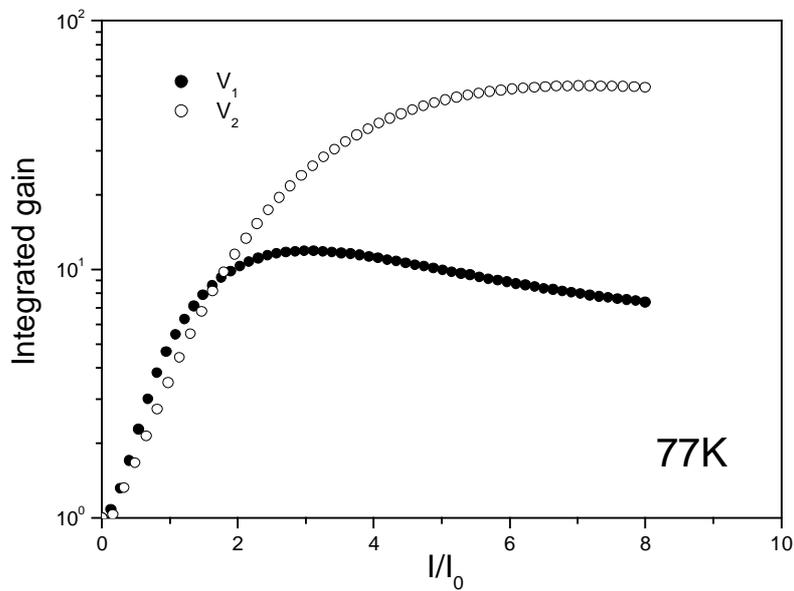

Figure 3b

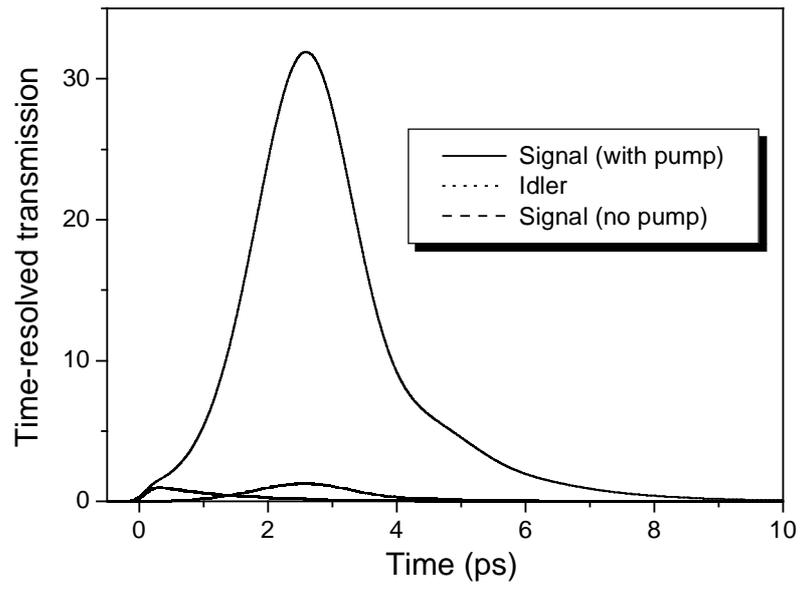

Figure 4